\documentclass[11pt]{article}
\pdfoutput=1

\usepackage[arrowdel]{physics}
\usepackage{lmodern}
\usepackage{siunitx}
\usepackage{amsmath}
\usepackage{amssymb}
\usepackage{textcomp}
\usepackage{graphicx}
\usepackage[caption=false]{subfig}
\usepackage{tikz}
\usepackage{mhchem}
\usepackage{tipa}
\usepackage{hyperref}
\usepackage[capitalise]{cleveref}
\usepackage[margin=0.75in]{geometry}

\usetikzlibrary{patterns}

\Crefname{equation}{Eq.}{Eqs.}
\Crefname{figure}{Fig.}{Figs.}
\Crefname{tabular}{Tab.}{Tabs.}
\crefdefaultlabelformat{#2#1#3}
\creflabelformat{equation}{#2#1#3}

\hypersetup{hidelinks}

\newcommand{\etal}{\textit{et al.}\ }

\newcommand*{\hrx}{x}
\newcommand*{\hry}{y}
\newcommand*{\hrz}{z}
\newcommand*{\hra}{\alpha}
\newcommand*{\hrb}{\beta}

\newcommand*{\chimera}{\chi}
\newcommand*{\meta}{m}
\newcommand*{\ordparam}{r}
\newcommand*{\phase}{\phi}

\begin{document}
\markboth{H.\ M.\ Mitchell \etal}{Chimera States and Seizures in a Mouse Neuronal Model}

\title{Chimera States and Seizures in a Mouse Neuronal Model}

\author{Henry M. Mitchell \and Peter Sheridan Dodds \and J. Matthew Mahoney \and Christopher M. Danforth}

\maketitle

\begin{abstract}
  Chimera states---the coexistence of synchrony and asynchrony in a nonlocally-coupled network of identical oscillators---are often used as a model framework for epileptic seizures.
Here, we explore the dynamics of chimera states in a network of modified Hindmarsh-Rose neurons, configured to reflect the graph of the mesoscale mouse connectome.
Our model produces superficially epileptiform activity converging on persistent chimera states in a large region of a two-parameter space governing connections (a) between subcortices within a cortex and (b) between cortices.
Our findings contribute to a growing body of literature suggesting mathematical models can qualitatively reproduce epileptic seizure dynamics.

\end{abstract}


\section{Introduction}
\label{sec:intro}
\subsection{Chimera States}
\label{sec:intro_chimera}
\begin{figure*}[ht]
  \centering
  \includegraphics[width=\textwidth]{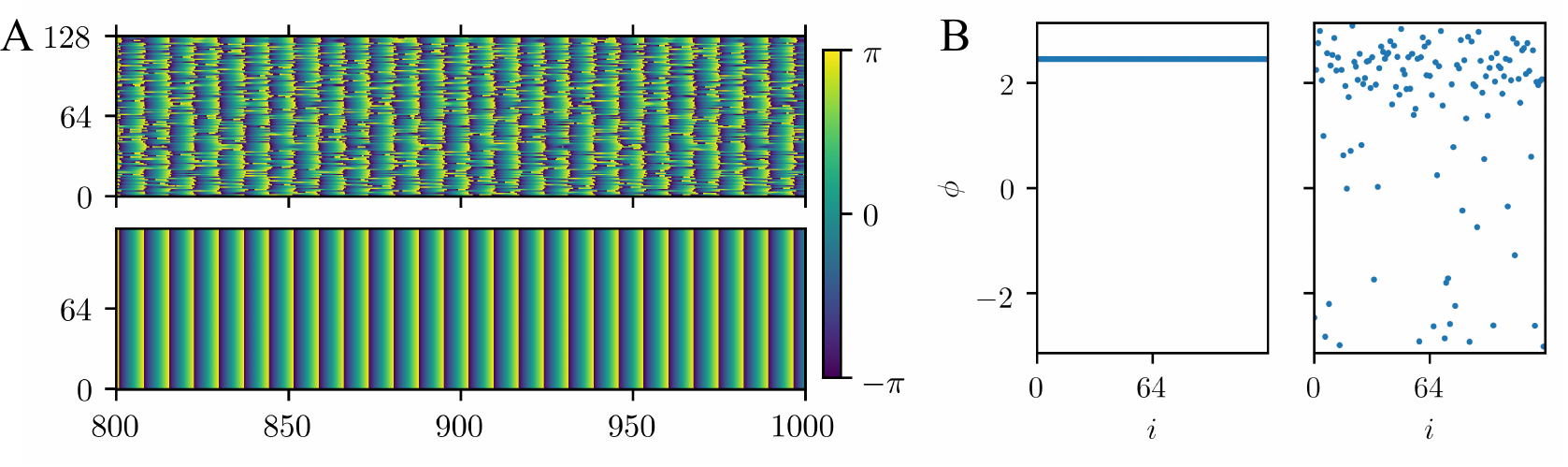}
  \caption[Abrams simulation]{A simulation of the Abrams model for two populations of 128 oscillators.
    We employed a 4th-order Runge-Kutta solver ($\dd{t} = 0.01$, $t_{\text{max}} = 1000$).
    A. Time series of the simulation for $t \in \pqty{800, 1000}$.
    B. Snapshot at $t \approx 800$.
  }
  \label{fig:abrams}
\end{figure*}

A well-studied example of complex behavior arising from simple mechanisms is the coexistence of synchrony and asynchrony within a system of identical coupled oscillators, a phenomenon known as a chimera state \cite{Kuramoto2002,Abrams2004}.
The existence of these chimera states is surprising, as they represent asymmetry within symmetric systems.
One of the simplest systems which produces chimera states is the Abrams model which consists of two populations of identical oscillators with a stronger coupling strength within the populations than between them \cite{Abrams2008}.
The system is described by the following:
\begin{equation}
  \label{eq:abrams}
  \dv{\theta_{i}^{\sigma}}{t}
  =
  \omega
  +
  \sum_{\sigma' = 1}^{2} \frac{K_{\sigma \sigma'}}{N_{\sigma'}} \sum_{j = 1}^{N_{\sigma'}} \sin(\theta_{j}^{\sigma'} - \theta_{i}^{\sigma} - \alpha),
\end{equation}
where
\begin{equation*}
  K
  =
  \bmqty{\mu & \nu \\ \nu & \mu}
  \qand
  \sigma \in \Bqty{1, 2}.
\end{equation*}
In this model, $\mu$ represents the intra-population strength, and $\nu$ represents the inter-population strength, with $\mu > \nu$.
Time can be scaled such that $\mu + \nu = 1$.
If $\mu - \nu$ is not too large, and $\alpha$ is not too much less than $\frac{\pi}{2}$, then this system can produce chimera states.
\Cref{fig:abrams} shows a simulation of the Abrams model on two populations of 128 oscillators.

An analogous system has recently been analyzed in the physical world \cite{Martens2013}.
Two swinging platforms were coupled together with springs of variable spring constant $\kappa$, and 15 metronomes---all tuned to the same frequency---were placed on each platform.
The metronomes on the same platform are coupled through the motion of the swing, which heavily influences the motion of the metronomes, represented in the Abrams model by $\mu$.
The metronomes on opposite platforms are coupled through the springs, which is a much weaker interaction, represented in the Abrams model by $\nu$.
For a wide range of values of $\kappa$, all of the metronomes on one platform would synchronize, while the metronomes on other platform would remain asynchronous.

While chimera states may present themselves obviously when observed in a plot or the physical world, they can be harder to pin down analytically.
In order to do so, we will investigate a system of $M$ communities of nonlocally-coupled oscillators, and we sample their phases at times $t \in \bqty{1, \ldots, T}$.
A useful pair of measures for detecting the presence of a chimera state are the chimera-like index $\chimera$ and the metastability index $\meta$ \cite{Shanahan2010,Hizanidis2016}.
To develop these two measures, we start with the order parameter $\ordparam_{c}(t) = \abs{\expval{e^{i \phase_{k}(t)}}_{k \in C}}$, where $\phase_{k}$ is the phase of oscillator $k$, and $\ev{f}_{k \in C}$ is the average of $f$ over all $k$ in community $C$.
The order parameter $\ordparam$ indicates the instantaneous synchrony of a community (how similar the phases of the oscillators are to the others in $C$), and not its overall coherence (how similar the trajectories of the oscillators are).
From this, we define the two measures:
\begin{align}
  \label{eq:chimera}
  \chimera
  &=
    7 \times \expval{\sigma_{\text{chi}}}_{T}, \\
  \label{eq:metastability}
  \meta
  &=
    12 \times \expval{\sigma_{\text{met}}}_{C},
\end{align}
where
\begin{equation}
  \sigma_{\text{chi}}(t)
  =
  \frac{1}{M - 1} \sum_{c \in C}\pqty{\ordparam_{c}(t) - \expval{\ordparam_{c}}_{C}}^{2},
\end{equation}
and
\begin{equation}
  \sigma_{\text{met}}(c)
  =
  \frac{1}{T - 1} \sum_{t \leq T}\pqty{\ordparam_{c}(t) - \expval{\ordparam_{c}}_{T}}^{2}.
\end{equation}
To put this into words, the chimera-like index $\chimera$ is the average over time of the variance of the order parameter across communities, while the metastability index $\meta$ is the average across communities of the variance of the order parameter within a given community over time,

The normalization constants follow from the indices' maximum possible values \cite{Shanahan2010}.
If a community spends equal time in a maximally chimeric state and a minimally chimeric state, then its chimera-like index will be at its maximum\footnote{While it is possible for half of a system's communities to be synchronous and the other half asynchronous for all times (resulting in a chimera-like index of $\frac{2}{7}$), this is transient due to the effects of metastability \cite{Shanahan2010}.  Therefore, we will ignore this case.}: $\chimera_{\text{max}} = \frac{1}{7}$.
If a community $c$ spends equal time in all stages of synchronization (i.e., the phase parameter of $c$ is uniformly distributed), then $\sigma_{\text{met}}(c)$ is at its maximum, which is the variance of the uniform distribution: $\meta_{\text{max}} = \frac{1}{12}$.


\subsection{Seizures}
\label{sec:intro_seizures}
Chimera states have been observed in many other systems, whether they be purely mathematical, biological, electrical, or mechanical \cite{Shanahan2010,Abrams2004,Andrzejak2016,Hizanidis2016,Kuramoto2002,Martens2013,Panaggio2015,Santos2015,Santos2017,Kruk2018,Xie2014}.
One of the most common ways that chimera states are discussed is in regards to seizures.

As is often the case with emergent phenomena, it is wildly impractical to simulate the collective behavior of a brain by simulating its constituent neurons.
Since the human brain has approximately $10^{11}$ neurons with $10^{14}$ synapses, direct simulation is too computationally intensive.
To better understand the dynamics of large portions of the brain, many researchers have turned to the techniques of thermal and statistical physics \cite{Breakspear2017},
resulting in \textit{neural field models} and \textit{neural mass networks}.
The first treats the brain as a continuous sheet of cortex, within which activity obeys wave equations.
The second represents the brain as a discrete graph of cortices, or a network of coupled oscillators.
The network used for the coupling of the oscillators is determined by the brain's connectivity matrix, or connectome.
An example of a neural mass network model is the modified Hindmarsh-Rose model [\cref{eq:hr_x,eq:hr_y,eq:hr_z}], which we discuss later.

One of the benefits of a neural mass network model is that its outputs are similar to those of an electroencephalograph, or EEG.
The EEG is a device used to record the electrical activity of the brain.
Electrodes are placed in specific areas on the scalp, and then changes in voltage are measured from neural masses beneath the skull.
Much of the signal is distorted and attenuated by the bone and tissue between the brain and the electrodes, which act like resistors and capacitors.
This means that, while the membrane voltage of the neuron changes by millivolts, the EEG reads a signal in the microvolt scale \cite{Kandel2013}.
The EEG also has relatively low spatial and temporal resolution (16 electrodes for the whole brain, and a sampling rate of \SI{33}{\ms}).
However, when properly treated, neural mass models make for effective predictors of the output from EEGs \cite{Taylor2012,Leistritz2007}.
This is useful, as EEGs are the main tool used to detect and categorize seizures.

\subsubsection{Seizure \AE tiology}
\label{sec:intro_seizures_aetiology}
Researchers define seizures as abnormal, excessive, or overly-synchronized neural activity \cite{Kandel2013,Baier2012}.
It is important to distinguish between seizures and epilepsy, as the two are often conflated.
Seizures are an acute event, whereas epilepsy is a chronic condition of repeated seizures.
While classification schemes vary, all center around the division between generalized and focal seizures.

The focus of the present work is focal seizures, which start in one part of the brain (the seizure focus).
They are typically preceded by auras such as a sense of fear, or hearing music, and often manifest as clonic movement of the extremities.
In many cases, they secondarily generalize, spreading to the entire brain.
This can make focal seizures and primary generalized seizures hard to distinguish, as a focal seizure can generalize rapidly after a brief aura.
This can lead to misdiagnoses and improper treatments.



\section{Literature Review}
\label{sec:lit_review}
Chimera states in brain models have often been linked loosely to
unihemispheric sleep,
seizures,
and other brain behaviors \cite{Abrams2008,Panaggio2015,Martens2013,Abrams2004,Shanahan2010,Hohlein2019,Bansal2019,Chouzouris2018}.
In most cases,
these connections are made in off-hand remarks to introduce the concept of a chimera state,
but serious connections between these phenomena are rarely drawn.
However,
there are some notable cases of investigations of chimera states in brains.

A example of a chimera state being investigated in neural models was an exploration of chimera states on a network of Hindmarsh-Rose neurons (\cref{eq:hr_x,eq:hr_y,eq:hr_z}) \cite{Santos2017}.
This model was simulated on the connectome of a cat.
Parameter space for the two connection strengths $\hra$ and $\hrb$ was explored.
Chimera states are most prevalent for low values of $\hrb$, the inter-cortex connection strength \cite{Santos2017}.
This is unsurprising.
If the inter-cortex connection strength is too high as compared to the intra-cortex connection strength, the coupling acts globally instead of nonlocally.
This means that each cortex has less holding it together than pulling it apart, allowing the system to descend into asynchrony.

Further, with increasing input current $I_{j}$ (and increasing noise in the input current), chimera states give way to incoherence.
This also intuitively makes sense.
As the input current increases, its significance relative to the coupling also increases.
Thus, the oscillators have no reason to synchronize.
And, of course, adding noise will simply amplify the effect \cite{Santos2017}.



\section{Methods}
\label{sec:methods}
\subsection{Model}
\label{sec:methods_model}
The model we used was a modified Hindmarsh-Rose neural model\footnote{The modification is to add in the coupling, turning it into a network model instead of a single neuron.
  The brain has different connection strengths and types within cortices than it does between cortices, which is why the network is separated into intra- and intercortical connections in \cref{eq:hr_x}.} taken from \cite{Santos2017}.
\begin{align}
  \begin{split}
  \label{eq:hr_x}
  \dot{\hrx}_{j}
  ={}&
    \hry_{j}
    -
    \hrx_{j}^{3}
    +
    b \hrx_{j}^{2}
    +
    I_{j}
    -
    \hrz_{j} \\
    & -
    \frac{\hra}{n'_{j}} \sum_{k = 1}^{N} G'_{j k} \Theta_{j}(\hrx_{k})
    -
    \frac{\hrb}{n''_{j}} \sum_{k = 1}^{N} G''_{j k} \Theta_{j}(\hrx_{k}),
    \end{split} \\
  \label{eq:hr_y}
  \dot{\hry}_{j}
  ={}&
    1
    -
    5 \hrx_{j}^{2}
    -
    \hry_{j}, \\
\end{align}
and
\begin{equation}
  \label{eq:hr_z}
  \dot{\hrz}_{j}
  ={}
    \mu \pqty{s \pqty{\hrx_{j} - \hrx_{\text{rest}}} - \hrz_{j}},
\end{equation}
where
\begin{equation}
  \label{eq:hr_sigmoid}
  \Theta_{j}(\hrx_{k})
  =
  \frac{\hrx_{j} - \hrx_{\text{rev}}}{1 + e^{-\lambda \pqty{\hrx_{k} - \theta}}}
\end{equation}
is the sigmoidal activation function.
This function helps the model better approximate the behavior of neural masses, as opposed to specific neurons.
\Cref{tab:hr_params} shows the values and meanings of the symbols in the model.

\begin{table}[ht]
  \centering
  \begin{tabular}{c | c | p{0.6\columnwidth}}
    Symbol & Value & Meaning \\ \hline
    $\hrx_{j}$ & --- & Membrane potential of the $j$th neural mass \\
    $\hry_{j}$ & --- & Associated with the fast processes \\
    $\hrz_{j}$ & --- & Associated with slow processes \\ \hline
    $b$ & 3.2 & Tunes the spiking frequency \\
    $I_{j}$ & 4.4 & External input current \\
    $\hrx_{\text{rev}}$ & 2 & Ambient reversal potential \\
    $\lambda$ & 10 & Activation function parameter \\
    $\theta$ & -0.25 & Activation function parameter \\
    $\mu$ & 0.01 & Time scale for variation of $z$ \\
    $s$ & 4 & Governs adaptation \\
    $\hrx_{\text{rest}}$ & -1.6 & Resting/equilibrium potential \\ \hline
    $\hra$ & Varied & Coupling strength within cortices \\
    $n_{j}'$ & See \cref{fig:primes} A & Number of connections within a cortex from the $j$th neuron \\
    $G_{j k}'$ & See \cref{fig:primes} C & Intra-cortical connection strength \\
    $\hrb$ & Varied & Coupling strength between cortices \\
    $n_{j}''$ & See \cref{fig:primes} B & Number of connections between cortices from the $j$th neuron \\
    $G_{j k}''$ & See \cref{fig:primes} D & Inter-cortical connection strength
  \end{tabular}
  \caption[Hindmarsh-Rose Parameters]{The list of parameters used in modeling the Hindmarsh-Rose network.}
  \label{tab:hr_params}
\end{table}

The measurable output of an EEG corresponds to the mean of the membrane potential within a cortex
(i.e., the observable values are $\expval{x_{j}}_{j \in C}$).

We chose this model due to the intelligibility of its parameters, as well as its proven ability to exhibit chimera-like behavior as a neural mass model \cite{Santos2017}.
Additionally, the Hindmarsh-Rose model was not designed to emulate seizures, which provides further evidence for the assertion that chimeras may be a universal aspect of brain activity.

It is worth noting that one of the limitations of this model is the changing nature of intra- and inter-cortical connection strengths (corresponding to $\hra$ and $\hrb$) in the actual brain.
The strengths of connections and the amounts by which they are amplified vary in time.
However, they will be treated as constant, in order to present a view of parameter space.

Additionally, the model assumes a network of identical nodes.
This is not the case in the brain, as there is heterogeneity of neuronal behavior, even among neurons of similar types.
The model does allow some flexibility for creating differences between neural masses through the $I_{j}$ external input current term.
Changes to this term have been shown to disrupt chimera states \cite{Santos2017}, possibly due to intermediate levels of heterogeneity enabling optimal information encoding \cite{Baroni2014}.
However, a network of identical nodes provides a good qualitative picture of the parameter space, with room for refinement by the introduction of noise.


\subsection{Network}
\label{sec:methods_network}
We implemented the model on a mesoscale mouse connectome,
which comprises of 213 fine areas,
grouped into 13 coarse areas,
along with measured connection strengths between the subcortices \cite{Oh2014}.
We used these coarse areas as the sets $C$ (\cref{eq:chimera,eq:metastability}) for chimera and metastability analyses.
We reduced the connection strengths to those with sufficient certainty ($p < 0.01$), and segmented as follows:
\begin{equation}
  \label{eq:mouse_segmentation}
  G_{j k}
  =
  \begin{cases}
    0 \text{ if } O_{j k} < 10^{-4}, \\
    1 \text{ if } 10^{-4} \leq O_{j k} < 10^{-2}, \\
    2 \text{ if } 10^{-2} \leq O_{j k} < 1, \\
    3 \text{ if } 1 \leq O_{j k},
  \end{cases}
\end{equation}
where $O_{j k}$ is the raw connection strength provided by \cite{Oh2014}.
We performed this simplification to match the analysis of Santos \etal more closely \cite{Santos2017}.

\begin{figure*}[ht]
  \centering
  \includegraphics[width=\textwidth]{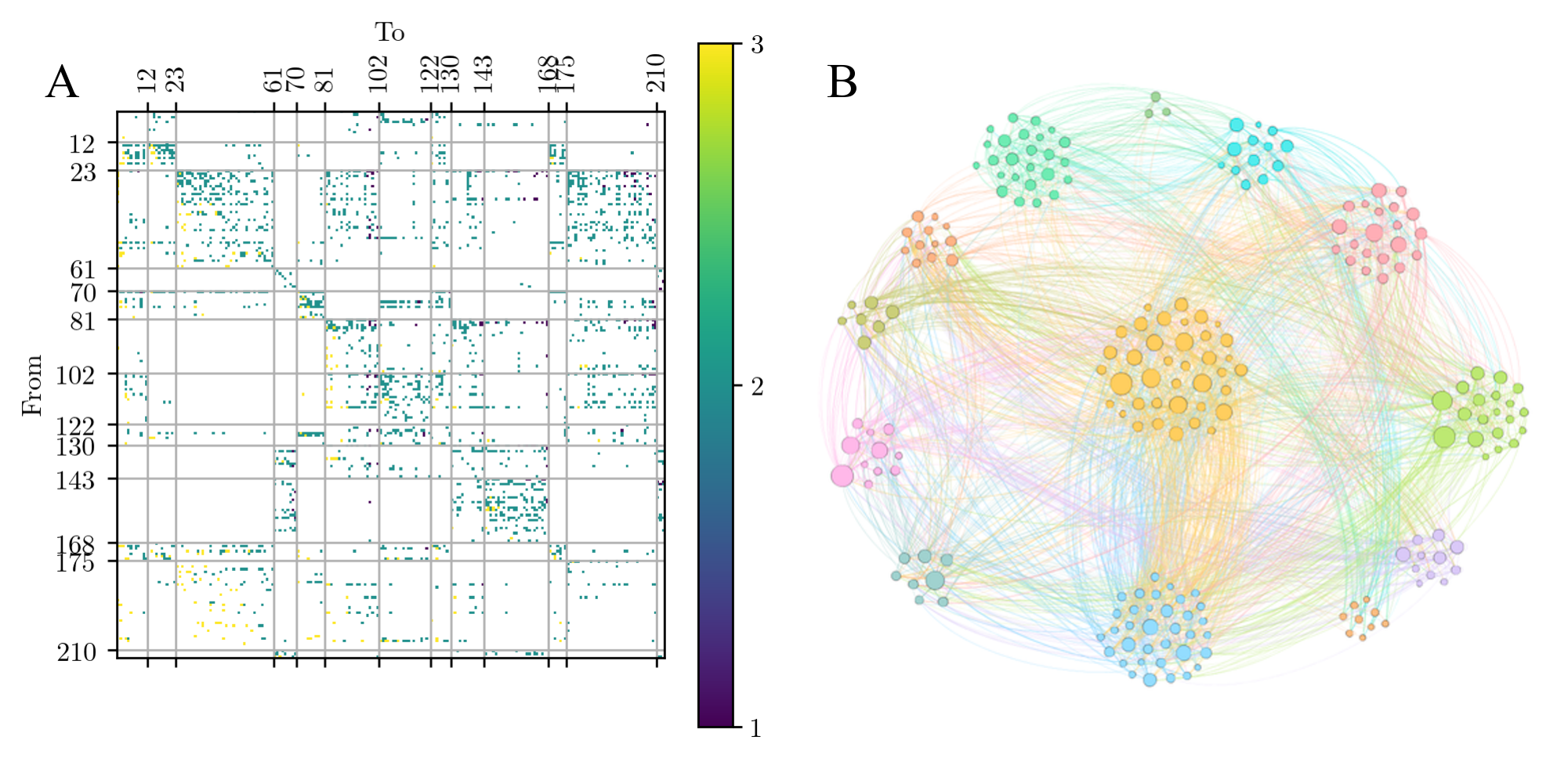}
  \caption[Mouse connectome]{A. A matrix representation of the mouse connectome, with strengths as defined by \cref{eq:mouse_segmentation}.
    The cortices represented are, left to right (top to bottom),
    the striatum,
    the olfactory areas,
    the isocortex,
    the crebellar cortex,
    the hippocampal formation,
    the midbrain,
    the hypothalamus,
    the pallidum,
    the pons,
    the medulla,
    the cortical subplate,
    the thalamus,
    and the cerebellar nuclei.
    B. An embedding of the graph.
    Edge colors indicate the source location.
  }
  \label{fig:mouse_connectome}
\end{figure*}

\begin{figure}[ht]
  \centering
  \includegraphics[width=0.5\textwidth]{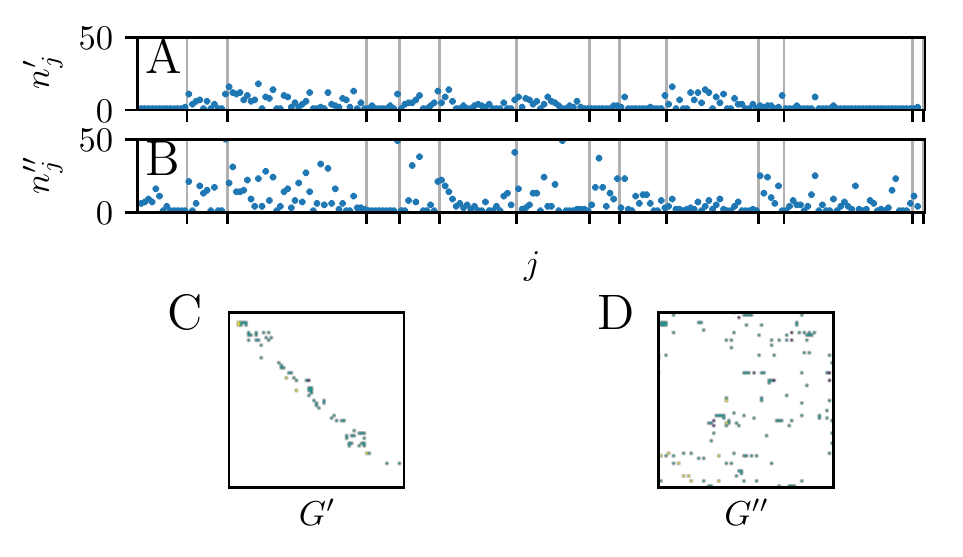}
  \caption[Network breakdown]{A breakdown of the network.
    A.\ \& B.\ show $n_{j}'$ and $n_{j}''$, effectively the number of nonzero elements in the $j$th row of $G'$ and $G''$, respectively.
    C.\ \& D.\ show $G'$ and $G''$, which are $G$ (\cref{fig:mouse_connectome} A) only within and between cortices, respectively.
  }
  \label{fig:primes}
\end{figure}

We show $G$ in \cref{fig:mouse_connectome}, and break it down into its inter- and intra-connections in \cref{fig:primes}.
This brain network is a small-world network \cite{Oh2014},
a graph topology which lends itself well to the development of chimera states, as it facilitates nonlocal coupling \cite{Hizanidis2016}.

Another benefit to this network is that it is comparatively accurate and complete.
Given the complexity of brains, creating an accurate structural or functional connectome is extremely difficult.
It has yet to be done to a large-scale extent in humans, and was only recently done in mice.
Moreover, as mice are common analogues for humans in laboratory settings, the mouse seemed a fitting ``guinea pig'' for the creation of chimera states.

\begin{figure}[ht]
  \centering
  \includegraphics[width=0.5\textwidth]{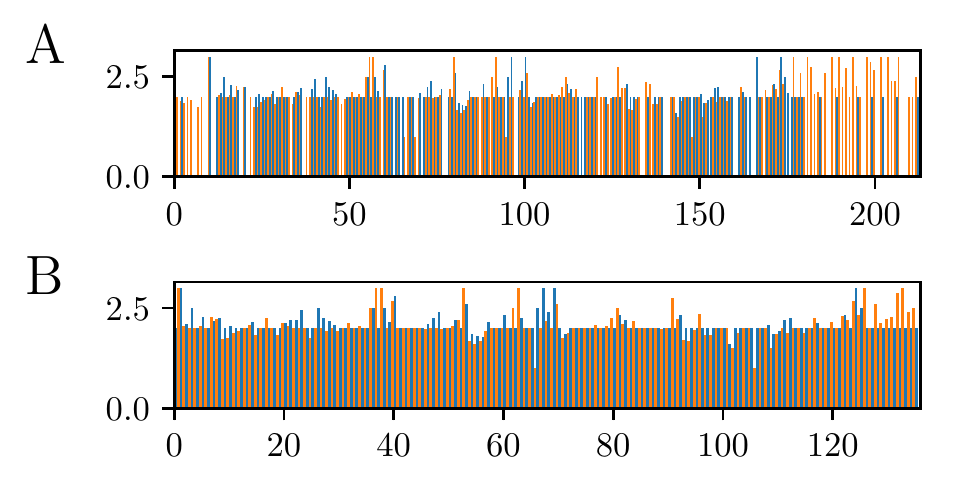}
  \caption[Average strengths]{The average connection strengths for each neuron $j$, within cortices (blue) and between them (orange).
    A.\ All of the subcortices.
    B.\ All of the subcortices for which neither intra- nor inter-cortical average strength was 0.
  }
  \label{fig:average_strengths}
\end{figure}


\subsection{Implementation}
\label{sec:methods_implementation}
We coded the modified Hindmarsh-Rose model using Python (Python version 3.7.0, NumPy version 1.15.2, Pandas version 0.23.4, SciPy version 1.1.0), and integrated using a 4th-order Runge-Kutta with variable step size\footnote{Step size was determined by SciPy's internal algorithms, but was limited to a maximum of 0.01.} $\dd{t} < 0.01$.
We verified the code by reproducing the results of \cite{Santos2017}.
We ran the model for a time period of $T_{\text{sim}} = \bqty{-1000, 5000}$, where only times $T = \bqty{0, 4000}$ were saved.
We threw away the times $\bqty{-1000, 0}$ to eliminate transients.
The chimeras were extremely unlikely to be eliminated on such a time scale, due to the size of the network \cite{Wolfrum2011}.
We calculated the times $\bqty{4000, 5000}$ to facilitate analysis of the phase.

We computed the phase of the $j$th neuron in the resulting waveform as
\begin{equation}
  \label{eq:hr_phase}
  \phase_{j}(t)
  =
  2 \pi \times \frac{t - t_{i}}{t_{i + 1} - t_{i}},
\end{equation}
where $t_{i}$ is the time at which the $j$th neuron fires ($x_{j}$ crosses 0 in a positive direction) for the $i$th time\footnote{This is a similar measure for the phase as was used in \cite{Santos2017}, but allows for easier discrimination between physical and aphysical parameter sets.
  It is modified to keep $\phase_{j} \in \pqty{0, 2\pi}$ and to eliminate ambiguity about the meanings of the subscripts.
}.
In order for this calculation to be possible for all values in $T$, it was necessary to have each neuron fire at least once after $T$ had finished (i.e., there has to be some $t_{i + 1} \notin T$ in order to calculate the phase for times $t_{i} \leq t \leq t_{\text{max}} = 4000$).
The calculated time range went so far beyond $t_{\text{max}}$ so that any extremely slow-firing neurons were allowed to do so, to ensure that as much of parameter space was in the physical region (\cref{sec:results_aphysical}).
We then used phase to find the chimera and metastability indices of the result using \cref{eq:chimera} and \cref{eq:metastability} respectively.

We repeated this process for various parameter sweeps of $\hra \times \hrb$, summarized in \cref{tab:parameter_sweeps}.
Note that the step in each strength $i \in \Bqty{\hra, \hrb}$ is $\Delta i = \frac{i_{\text{max}}}{n_{i} - 1}$, due to the fact that the ranges are inclusive of both endpoints.
\begin{table}[ht]
  \centering
  \begin{tabular}{c | c | c}
    $\hra_{\text{max}}$, $\hrb_{\text{max}}$ & $\Delta \hra$ ($n_{\hra}$), $\Delta \hrb$ ($n_{\hrb}$) & Figure \\ \hline
    1.6, 0.4 & 0.0203 (80), 0.0211 (20) & --- \\
    3.2, 0.8 & 0.0405 (80), 0.0205 (40) & --- \\
    0.2, 0.1 & 0.00253 (80), 0.00256 (40) & \cref{fig:zoom} \\
    0.9, 0.9 & 0.0101 (80), 0.0101 (80) & --- \\
    1.0, 1.0 & 0.0101 (100), 0.0101 (100) & \cref{fig:aphysical_chimera}
  \end{tabular}
  \caption[Parameter sweeps]{The sweeps we used in evaluating the effects of $\hra$ and $\hrb$ on the chimera and metastability indices.
    All parameter sweeps started at $(\hra, \hrb) = (0, 0)$.
    We performed the $\bqty{0, 1} \times \bqty{0, 1}$ sweep 10 times, and averaged the resulting chimera-like indices.
  }
  \label{tab:parameter_sweeps}
\end{table}

Initial conditions were drawn from uniform distributions of $x_{j} \in \bqty{-2, 2}$, $y_{j} \in \bqty{0, 0.2}$, $z_{j} \in \bqty{0, 0.2}$.
We performed all simulations on the \href{https://www.uvm.edu/vacc}{Vermont Advanced Computing Core}, and is \href{https://github.com/henmitch/chimera-2019}{available online}\footnote{\href{https://github.com/henmitch/chimera-2019}{https://github.com/henmitch/chimera-2019}}.



\section{Results}
\label{sec:results}
We investigate three aspects of the model's output.
First, we compare the output from the model to real-world data, on a qualitative level.
We then discuss the region of parameter space for which the model produces aphysical results.
Finally, we draw connections between chimera states in the model and their physiological analogues.
\subsection{Model Quality}
\label{sec:results_model}
\begin{figure}[ht]
  \centering
  \includegraphics[width=0.5\textwidth]{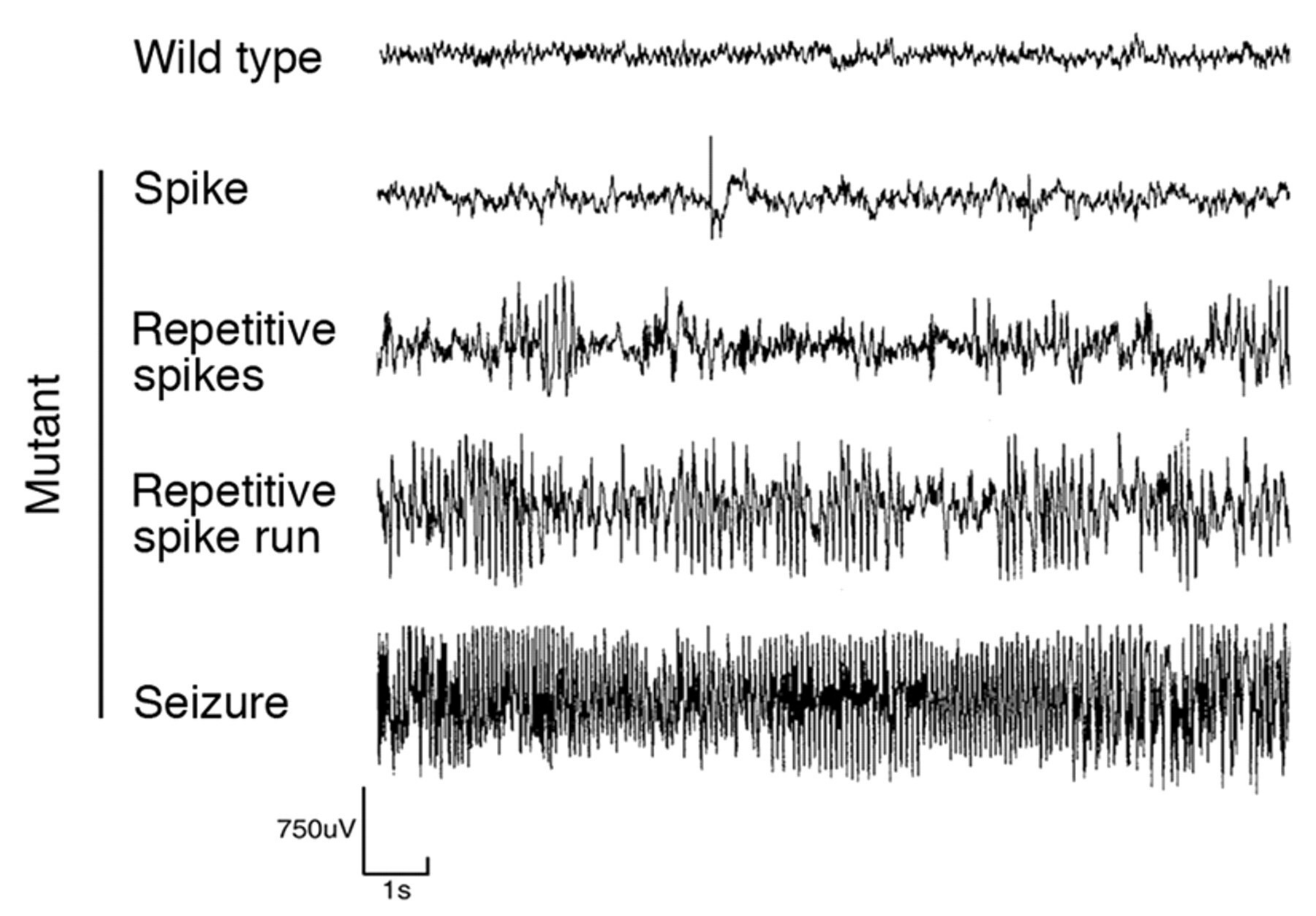}
  \caption[Typical EEG trace]{A typical EEG trace.
    The first row (``Wild type'') shows a normal awake adult mouse EEG trace.
    The other four rows (``Mutant'') show typical abnormal/epileptiform activity.
    Taken from \cite{Ljungberg2009}.
  }
  \label{fig:eeg}
\end{figure}

It is worthwhile to first discuss the quality of the model used, and its relationship to reality.
\Cref{fig:eeg} shows several types of behavior one can expect on an EEG trace.
Healthy brain behavior presents as low-amplitude oscillations on an EEG, as the asynchrony leads the firings of individual neurons to cancel each other out.
Seizures and seizure-like activity present as higher-amplitude oscillations, as the synchrony decreases the variance between neurons, making the mean closer to the behavior of each neuron.
One of the main challenges of simulating seizures is that only trained experts can truly identify seizures; as yet, there is no mathematical technique for identifying seizures \cite{Kandel2013}.

\begin{figure*}[ht]
  \centering
  \includegraphics[width=\textwidth]{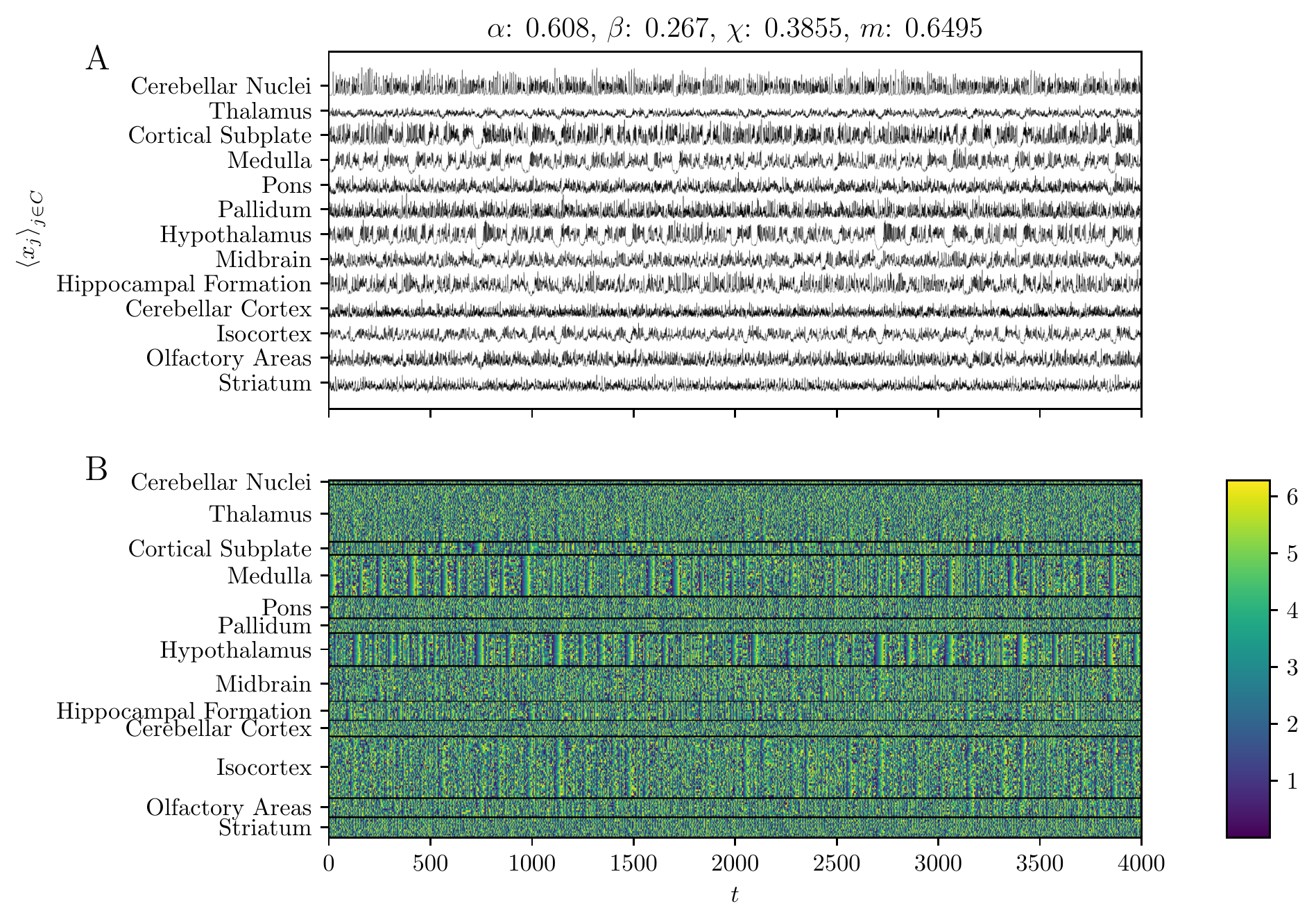}
  \caption[Typical simulation]{A typical run of the Hindmarsh-Rose simulation.
    A. The mean membrane potential within each cortex.
    B. The phase $\phase$ of the entire timeseries for a simulation of the Hindmarsh-Rose network.
  }
  \label{fig:608_267}
\end{figure*}

The highly chimeric portion of the landscape appears to be mostly below the $\beta = \alpha$ line.

However, one can indicate whether a model resembles epileptiform activity on a qualitative level.
\Cref{fig:608_267} shows the results of a simulation of the Hindmarsh-Rose network for $\pqty{\hra, \hrb} = \pqty{0.608, 0.267}$.
Qualitatively, in our simulation,
the thalamus, the pons, and the striatum look like the wild type EEG;
the cerebellar cortex shows some spiking behavior,
as well as some spike runs;
the medulla and the hypothalamus look to be in repetitive spike runs;
and the cortical subplate seems to be exhibiting seizure-like behavior over some time periods.
This shows that the behavior visible on an EEG can be approximately reproduced in this model.

While obtaining a higher neuronal resolution from the EEG is not possible due to the nature of the method,
we can see the neural dynamics of the model (\cref{fig:608_267} B).
It is evident that the thalamus, the pons, and the striatum are each highly asynchronous, which corresponds to their wild-type presentation.

Both presentation methods have their benefits, as \cref{fig:608_267} A looks similar to an EEG (and therefore lends itself well to the comparison), where \cref{fig:608_267} B allows us to view the individual subcortical behaviors leading to the net dynamics.

\subsection{Aphysical Region}
\label{sec:results_aphysical}
\begin{figure}[ht]
  \centering
  \includegraphics[width=0.5\textwidth]{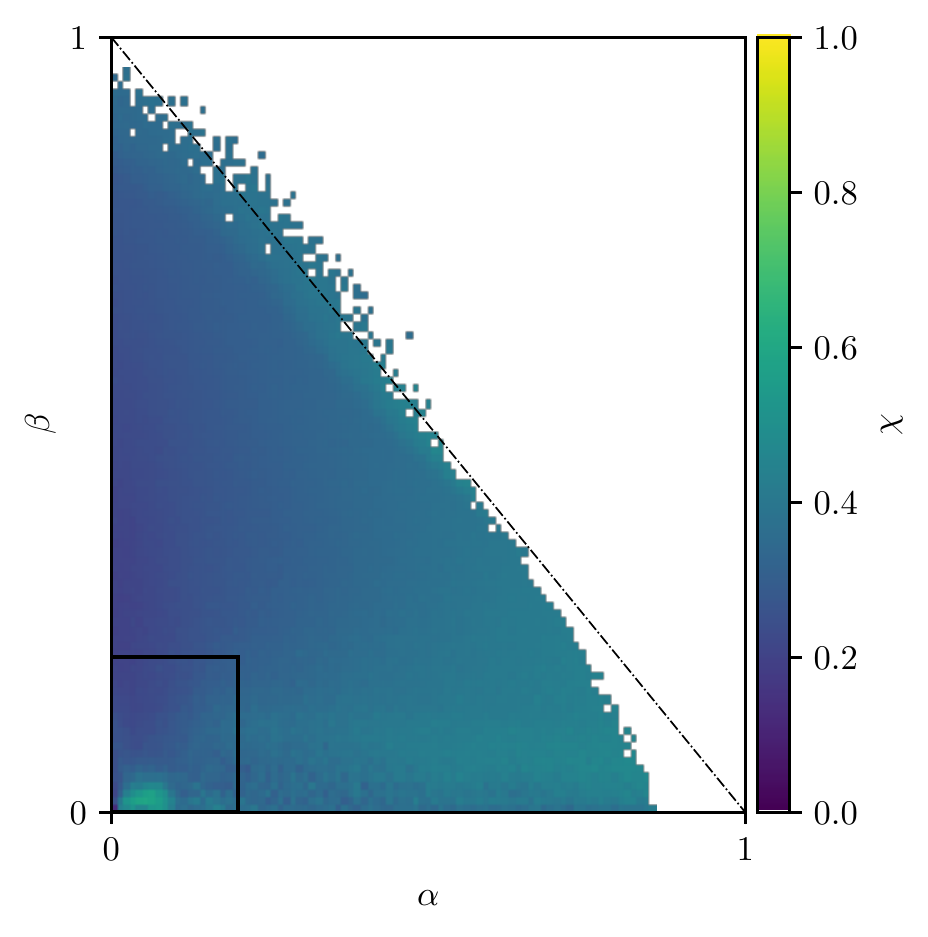}
  \caption[Chimera-like index landscape]{
    The chimera-like landscape of parameter space on $\pqty{\hra, \hrb} \in \pqty{0, 1.0} \times \pqty{0, 1.0}$.
    The aphysical region of the model is shown in white.
    The black rectangle in the bottom left corner indicates the region of parameter space shown in \cref{fig:zoom}.
    The dashed line has a slope of $-1$, to serve as a guide for \cref{sec:results_chimera}.
    The chimera-like index (defined in \cref{eq:chimera}) is normalized to 1, as usual.
  }
  \label{fig:aphysical_chimera}
\end{figure}

We choose not to include figures for the first two sweeps of \cref{tab:parameter_sweeps} because a large portion of parameter space leads to an aphysical model.
Specifically, for certain value pairs of $\pqty{\hra, \hrb}$, certain neurons never fired (increased past 1).
Despite the drastically increased time of evaluation (see \cref{sec:methods_implementation}), a vast swath of parameter space gave nonsense results (the white shown in \cref{fig:aphysical_chimera}).
The boundary between physical and aphysical appears to be linear, with a negative slope.
This means that $\hra$ can range further when $\hrb$ is low,
and vice versa.
This makes sense, as increased $\hra$ and $\hrb$ influence the model in the same way (increasing the coupling and decreasing $\dot{\hrx}_{j}$).

Furthermore, the slope of the boundary is greater than $-1$, which means that $\hra$ has an greater influence on the physicality of the model.
This is also reasonable, but for slightly less self-evident reasons.
To explain why, we must look specifically at the coupling term from \cref{eq:hr_x}:
\[
  -\frac{\hra}{n'_{j}} \sum_{k = 1}^{N} G'_{j k} \Theta_{j}\pqty{\hrx_{k}}
  -
  \frac{\hrb}{n''_{j}} \sum_{k = 1}^{N} G''_{j k} \Theta_{j}\pqty{\hrx_{k}}.
\]
This coupling will, in fact, be positive, as $\Theta_{j}\pqty{\hrx_{k}} < 0$ if $\hrx_{j} < 2$, which is true almost all of the time.
This means that, as $\hra$ and $\hrb$ increase, so does the overall coupling strength.
So, there is some threshold $K$ for which the overall coupling is too strong if, for some $j$,
\begin{equation}
  \label{eq:coupling_inequality}
  \frac{\hra}{n'_{j}} \sum_{k = 1}^{N} G'_{j k} \abs{\Theta_{j}\pqty{\hrx_{k}}}
  +
  \frac{\hrb}{n''_{j}} \sum_{k = 1}^{N} G''_{j k} \abs{\Theta_{j}\pqty{\hrx_{k}}}
  >
  K_{j}.
\end{equation}
In order for $\hra$ to influence the coupling's proximity to $K$ more than $\hrb$ does, there must exist some $j$ such that
$\frac{1}{n'_{j}} \sum G'_{j k} \abs{\Theta_{j}\pqty{\hrx_{k}}}
>
\frac{1}{n''_{j}} \sum G''_{j k} \abs{\Theta_{j}\pqty{\hrx_{k}}}$.
Seeing as $\overline{g}'_{j} = \frac{1}{n'_{j}} \sum G'_{j k}$ and $\overline{g}''_{j} = \frac{1}{n''_{j}} \sum G''_{j k}$ are the average connection strength within and between cortices (shown in \cref{fig:average_strengths} A), these are simply a function of the topology of the graph.

It may look from \cref{fig:average_strengths} A like $\hrb$ should have more influence than $\hra$, as for most $j$, $\overline{g}''_{j} > \overline{g}'_{j}$.
However, for most of those cases, $\overline{g}'_{j_{0}} > \overline{g}''_{j_{0}} = 0$.
This means that, for those $j_{0}$, $\pdv{K_{j_{0}}}{\hra} = 0$.
So, those cases contribute to the value of the threshold, but do not influence the physicality's dependence on $\hra$ and $\hrb$.

If we remove the $j$ for which $0 \in \Bqty{\overline{g}'_{j}, \overline{g}''_{j}}$, we find that, on average, $\overline{g}'_{j} = 2.100$, slightly more than $\overline{g}''_{j} = 2.079$ (see \cref{fig:average_strengths} B).
This explains the slope of the boundary between the physical region and the aphysical region.


\subsection{Chimera states}
\label{sec:results_chimera}
We show the normalized chimera-like index of the entire physical region in \cref{fig:aphysical_chimera}.
Near the maximal edge of the physical region, the highest values of the chimera index appear to follow a slope of $-1$.
It is unsurprising that chimera states would be prevalent when the coupling is large (out near the boundary of the aphysical range).

\begin{figure}[ht]
  \centering
  \includegraphics[width=0.5\textwidth]{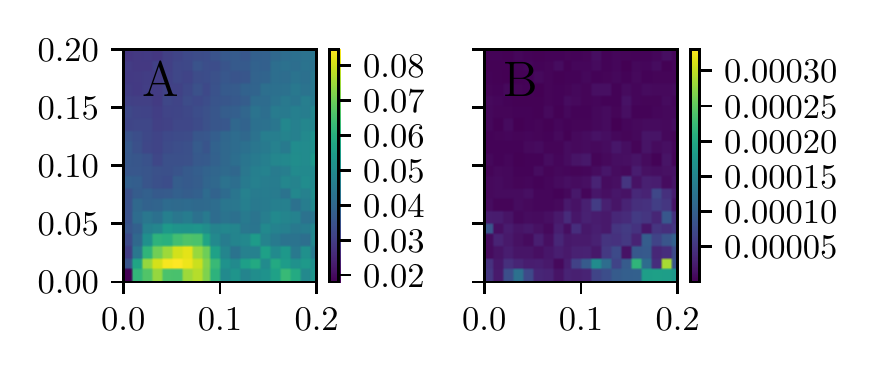}
  \caption[Zoomed landscape]{A.\ The chimera-like index $\chimera$ of runs with $(\hra, \hrb) \in (0, 0.2) \times (0, 0.2)$.
    As before, the chimera-like index is normalized to $\frac{1}{7}$.
    Note that the values of the index are much higher in this patch than in most of the rest of $(\hra, \hrb) \in (0, 1) \times (0, 1)$ (\cref{fig:aphysical_chimera}).
    B.\ The variance of the chimera-like index.
  }
  \label{fig:zoom}
\end{figure}

\begin{figure*}[ht]
  \centering
  \includegraphics[width=\textwidth]{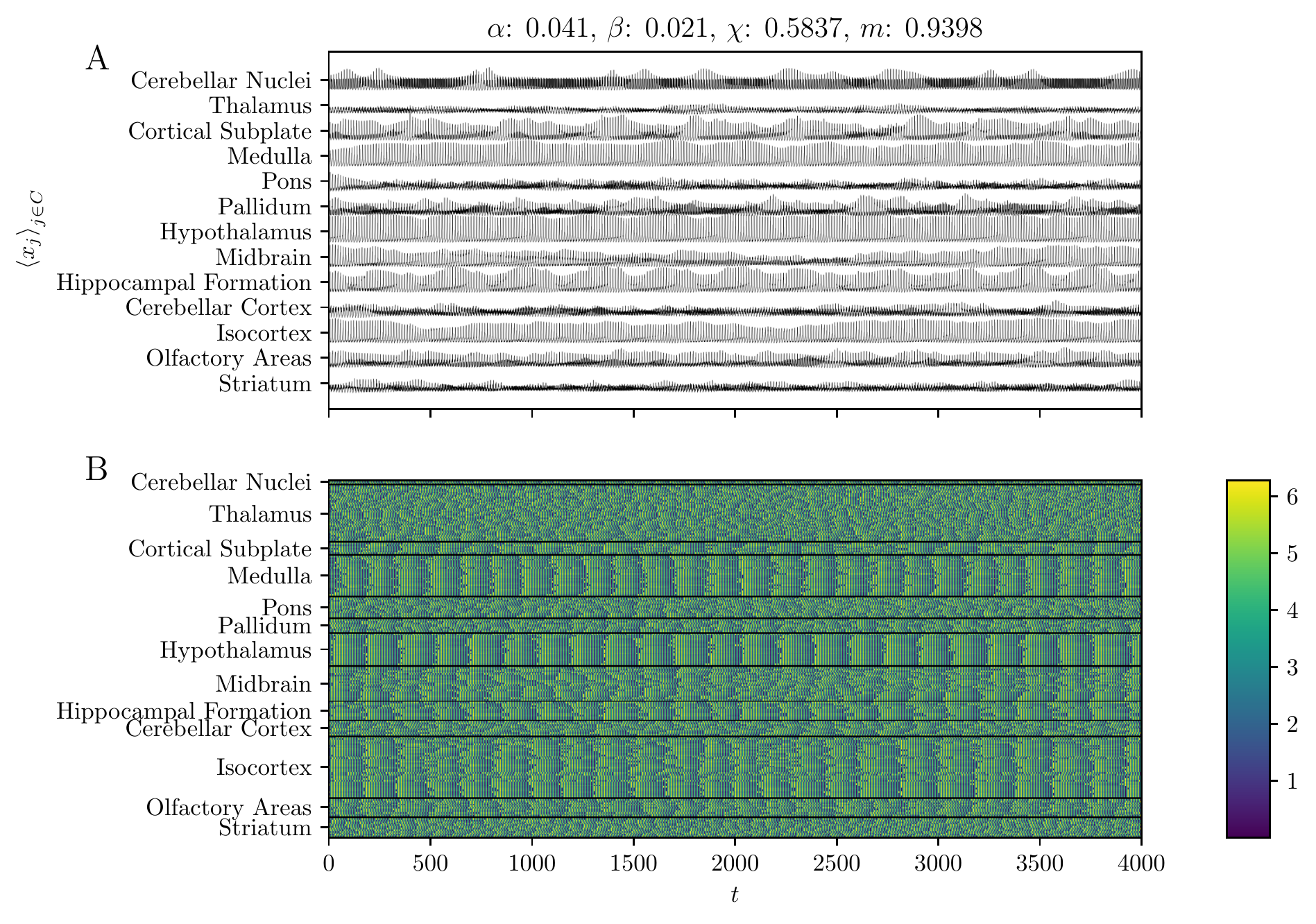}
  \caption[Highly chimeric simulation]{A run of the Hindmarsh-Rose simulation in the chimeric island.
    A. The mean membrane potential within each cortex.
    B. The phase $\phase$ of the entire timeseries for a simulation of the Hindmarsh-Rose network.
    Synchronization is most consistently evident in the medulla, the hypothalamus, and the isocortex.
  }
  \label{fig:041_021}
\end{figure*}

What is surprising, however, is the presence of the chimeric patch in the bottom left corner of \cref{fig:aphysical_chimera}, shown at a higher resolution in \cref{fig:zoom}.
Plotting the results of the simulations (\cref{fig:041_021}),
it is evident that this is not a calculation error, but is an actual feature of the parameter landscape.

The highly chimeric patch within the physical portion of the landscape appears to be mostly below the $\beta = \alpha$ line.
This is reasonable, as chimera states occur when coupling within groups is greater than coupling between groups.
A small portion of the chimeric patch lies above the $\beta = \alpha$ line, likely because the average strength between cortices is greater than the average strength within cortices (see \cref{fig:average_strengths} A).

The chimera-like index $\chimera$ greatly lessens at $\alpha \approx 0.1$.
A possible explanation for this comes from comparing the order of $\dot{\hrx}$ without the coupling terms, and the coupling terms themselves.
From our simulation, we find that $\dot{\hrx}$ without the coupling terms ranges roughly from -6 to 3.
The coupling terms each\footnote{Since the same can be said for both $\alpha$ and $\beta$, we will discuss only $\alpha$, with the understanding that $\beta$ could be substituted into the proceeding sentences.}
range from 0 to approximately $30 \alpha$.
This means that, when $\alpha > 0.1$, the coupling is at least of the same order as the sum of the rest of the terms in the equation.
This leads to a qualitative difference between the two states, which likely manifests itself as the less-chimeric states.


\section{Discussion}
\label{sec:conclusion}
The logical next step is to compare the simulated EEG traces to actual data collected from mice, with the aid of an epileptologist.
With further work in this direction, our research could potentially go from a mathematical curiosity to an applicable therapeutic and diagnostic tool.
However, the initial results are encouraging.
The majority of the simulated seizures showed excessive synchrony in areas such as the hypothalamus and the hippocampal formation, which are the most common origins of ictal activity \cite{Maguire2013,Avoli2007}.
The exact origin of focal seizures is poorly understood, and our implementation of the model makes it difficult to narrow in on the precise onset of chimera states.
This could be a future direction for research in this area.

Finding an instructive phase-space embedding of the Hindmarsh-Rose network would be challenging (seeing as it is a 639-dimensional system),
but would likely reveal potentially useful insights into the nature of the mechanisms underlying these systems.
The same could be said for Lyapunov analysis, as well as finding an informative way to create a bifurcation diagram and perform more in-depth bifurcation analysis.

Another way our work could be extended is by looking at chimera state collapse and its relationship to secondary seizure generalization.
However, it would be extremely computationally expensive, given the size of the system, and would therefore require some clever handiwork \cite{Wolfrum2011}.

Future work could also naturally be performed on better, more up-to-date connectomes \cite{Knox2019}.



\section*{Acknowledgments}
\label{sec:acknowledgements}
This work could not have been completed without help from Taras Lakoba and Sean Flynn.
The authors also thank Kameron Harris for helpful comments on the manuscript.

\begin{figure*}
  \centering
  \includegraphics{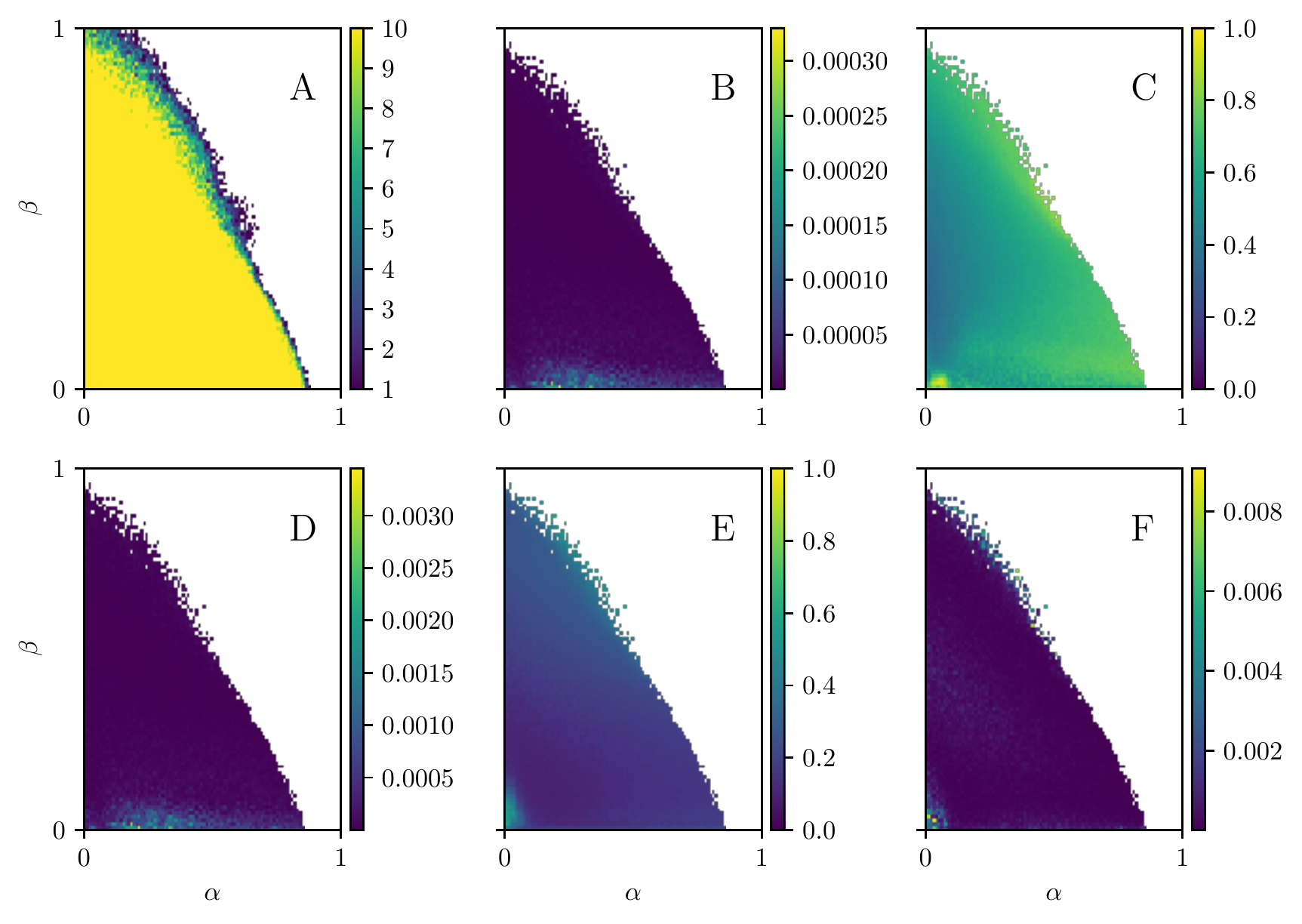}
  \caption[Additional figures]{Additional analyses of the parameter landscape.
    A.\ The number of runs (out of 10) which gave physical results.
    All parameter sets which are not yellow in A were excluded from the other analyses.
    B.\ The variance of the chimera-like index.
    C.\ The metastability index.
    D.\ The variance of the metastability index.
    E.\ The mean order parameter over time, averaged across runs.
    F.\ The variance between runs of the mean order parameter.
  }
  \label{fig:additional_figures}
\end{figure*}

\appendix{Additional Figures}
\label{sec:figures}
\cref{fig:additional_figures} shows some additional analyses of the parameter landscape.

\bibliography{ms.bib}

\end{document}